\documentclass{iau}
\usepackage{graphicx}

\title[Dwarf galaxies in nearby voids] 
{Observations of dwarfs in nearby voids: \\ implications for galaxy formation and evolution}

\author[Simon A. Pustilnik]   
{Simon A. Pustilnik$^1$}

\affiliation{$^1$Special Astrophysical Observatory of Russian Academy of
Sciences, \\ 369167, Nizhnij Arkhyz, Russia,
 \\ email: {\tt sap@sao.ru} }  %

\pubyear{2014}
\volume{308}  
\pagerange{119--126}
\jname{Title of your IAU Symposium}
\editors{R.~van~de~Weygaert, S.~Shandarin, E.~Saar and J.~Einasto, eds.}
\begin{document}

\maketitle

\begin{abstract}
The intermediate results of the ongoing study of deep samples of $\sim$200
galaxies residing in nearby voids, 
are presented. Their properties are probed via optical spectroscopy, $ugri$
surface photometry, and HI 21-cm line measurements, with emphasis on their
evolutionary status.
We derive directly the hydrogen mass M(HI), the ratio $M$(HI)/$L_{\rm B}$
and the evolutionary parameter gas-phase O/H. Their luminosities and
integrated colours are used to derive stellar mass $M_{*}$ and the second
evolutionary parameter -- gas mass-fraction ($f_{\rm g}$).
The colours of the outer parts, typically representative of the galaxy oldest
stellar population, are used to estimate
the upper limits on time since the beginning of the main SF episode.
We compare properties of void galaxies with
those of the similar late-type galaxies in denser environments. Most of void
galaxies show smaller O/H for their luminosity, in average by $\sim$30\%,
indicating slower evolution. Besides, the fraction of $\sim$10\% of the whole
void sample or $\sim$30\% of the least luminous void LSB dwarfs show the
oxygen deficiency by a factor of 2--5. The majority of
this group appear very gas-rich, with $f_{\rm g} \sim$(95--99)\%, while their
outer parts appear rather blue, indicating the time of onset of the main
star-formation episode of less than 1--4 Gyr. Such unevolved LSBD galaxies
appear not rare among the smallest void objects, but turned out practically
missed to date due to the strong observational selection effects. Our results
evidense for unusual evolutionary properties of
the sizable fraction of void galaxies, and thus, pose the task of better
modelling of dwarf galaxy formation and evolution in voids.
\end{abstract}

\firstsection 

\section{Introduction and Overview}

Studies of galaxies in low-density environments were tempting in the hope
to probe the basics of galaxy evolution in isolation. The modern concepts
suggest however
that even the most isolated objects are related to and influenced by the
baryon flows of adjacent filaments. Observational studies of galaxy samples
in voids, based on large deep surveys (SDSS, 2dRGS), were mostly limited by
large distant voids ($D \sim$ 100--200~Mpc). As a consequence, they probed
only the upper part of void galaxy luminosity function ($M_{\rm r} < -16$).
Only subtle or at most moderate
differences with wall galaxies  were found on their SFR and colours.

The complementary approach to study tens-hundred less luminous galaxies
in nearby voids was suggested in \cite{PaperI}. The first void
intrinisically faint galaxy sample (down to $M_{\rm B} = -11$) was drawn up
in the nearby Lynx-Cancer void ($D_{\rm centre} =$18~Mpc), currently
including over 100 objects. Several interesting findings on void galaxy
evolution (see below) evidense for importance of this direction and emphasize
the need of larger statistics and detailed studies of the least luminous
void galaxies.

\begin{figure}[b]
\begin{center}
 \includegraphics[angle=-90,width=12cm]{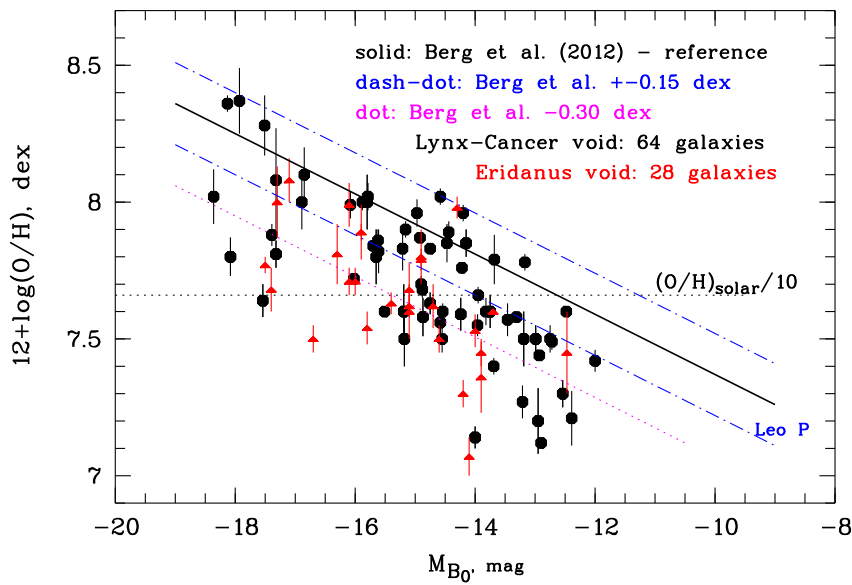}
 \caption{
O/H (with error bars) vs $M_{\rm B}$ relation for 92 galaxies in the
Lynx-Cancer (filled octogons) and Eridanus (filled triangles)
voids in respect of similar galaxies in denser environment (Berg et al.
2012, "reference"). The significant systematic O/H drop in void galaxies
is evident as well as the sizable fraction
of strong outliers (deficiency of O/H by factor of 2-5) (Pustilnik et al.
2011, 2014; Kniazev et al. 2014, in prep.)
}
   \label{fig1}
\end{center}
\end{figure}

\section{Ongoing project and intermediate results}

To significantly increase the number of faint void galaxies and conduct
more reliable statistical study of their properties, we work on the revision
of the sample of nearby voids and the sample of galaxies residing in them.
In particular, galaxies residing in Monoceros and Cetus voids and in the
equatorial part of Eridanus void (\cite{Fairall98}) are added to the current
sample of the Lynx-Cancer void.
Some of the important results on the evolutionary status of void galaxies
are illustrated below.

In Fig.~\ref{fig1} we summarise determinations of
gas-phase O/H in 92 galaxies residing in the Lynx-Cancer (filled octogons)
and Eridanus (filled triangles) voids
(\cite{void_OH}, Pustilnik et al. in prep., Kniazev et al. in prep.).
In the large fraction of studied spectra no [OIII]$\lambda$4363 line was
detected, and hence semi-empirical method by \cite{IT07} was used to determine
O/H. These O/H are shown vs absolute blue
magnitudes $M_{\rm B}$. For comparison we use the sample of similar galaxies
from the Local Volume for which the confident O/H, distances and $M_{\rm B}$
are known (\cite{Berg12}). Their linear regression of 12+$\log$(O/H) and
$M_{\rm B}$ is shown by solid line, while $\pm$1$\sigma$ rms scatter in O/H
(0.15 dex) are shown by dash-dotted lines. The substantial shift of the whole
void O/H  data below the `reference' Berg et al. relation is well seen.
Moreover, the sizable fraction of void galaxies shows the O/H deficiency of
more than by factor of two (up to five).

\begin{figure}[b]
\begin{center}
 \includegraphics[angle=-90,width=12cm]{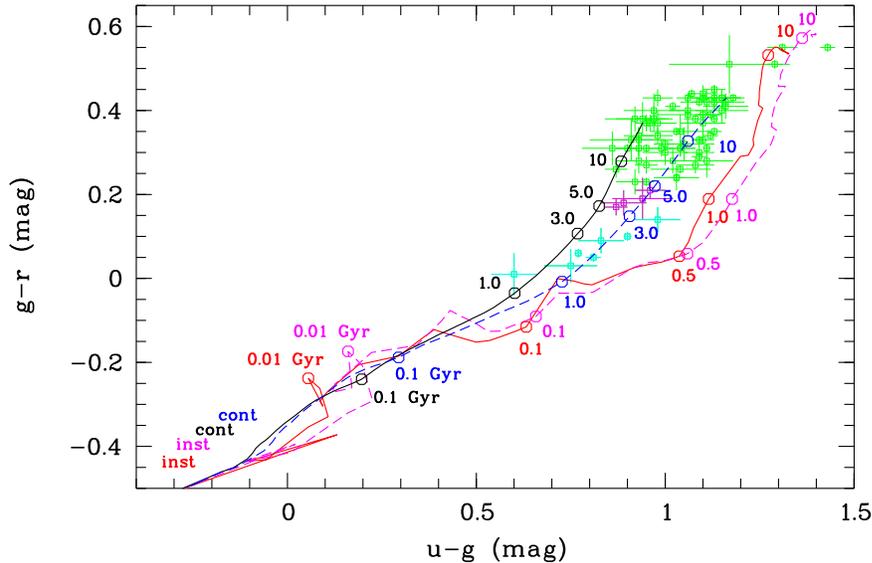}
 \caption{
Age indicators: $ugr$ colours of outer parts of 85 Lynx-Cancer void galaxies
superimposed on PEGASE2 evolutionary tracks. $\sim$15\% show retarded main
Star Formation, started only 1--5 Gyr ago (see  \cite{void_photo}). Solid
lines correspond to Salpeter IMF, while dashed lines - to Kroupa et al. IMF.
}
   \label{fig2}
\end{center}
\end{figure}

Another important result comes from mass photometric study of Lynx-Cancer
void galaxies based on the SDSS database (\cite{DR7}). In particular,
we determined $ugri$ colours of outer parts for 85 void galaxies and
compared them with PEGASE2 (\cite{pegase2})
evolutionary tracks for two extreme SF laws: instantaneous and continuous
with constant SFR (Fig.~\ref{fig2}). While the great
majority of void galaxies started their main star formation 7-14 Gyr ago,
for about 15\% of the sample we have clear indication of the retarded main
star  formation, commenced 1--5 Gyr ago.

\begin{figure}[b]
\begin{center}
 \includegraphics[angle=-0,width=6.5cm]{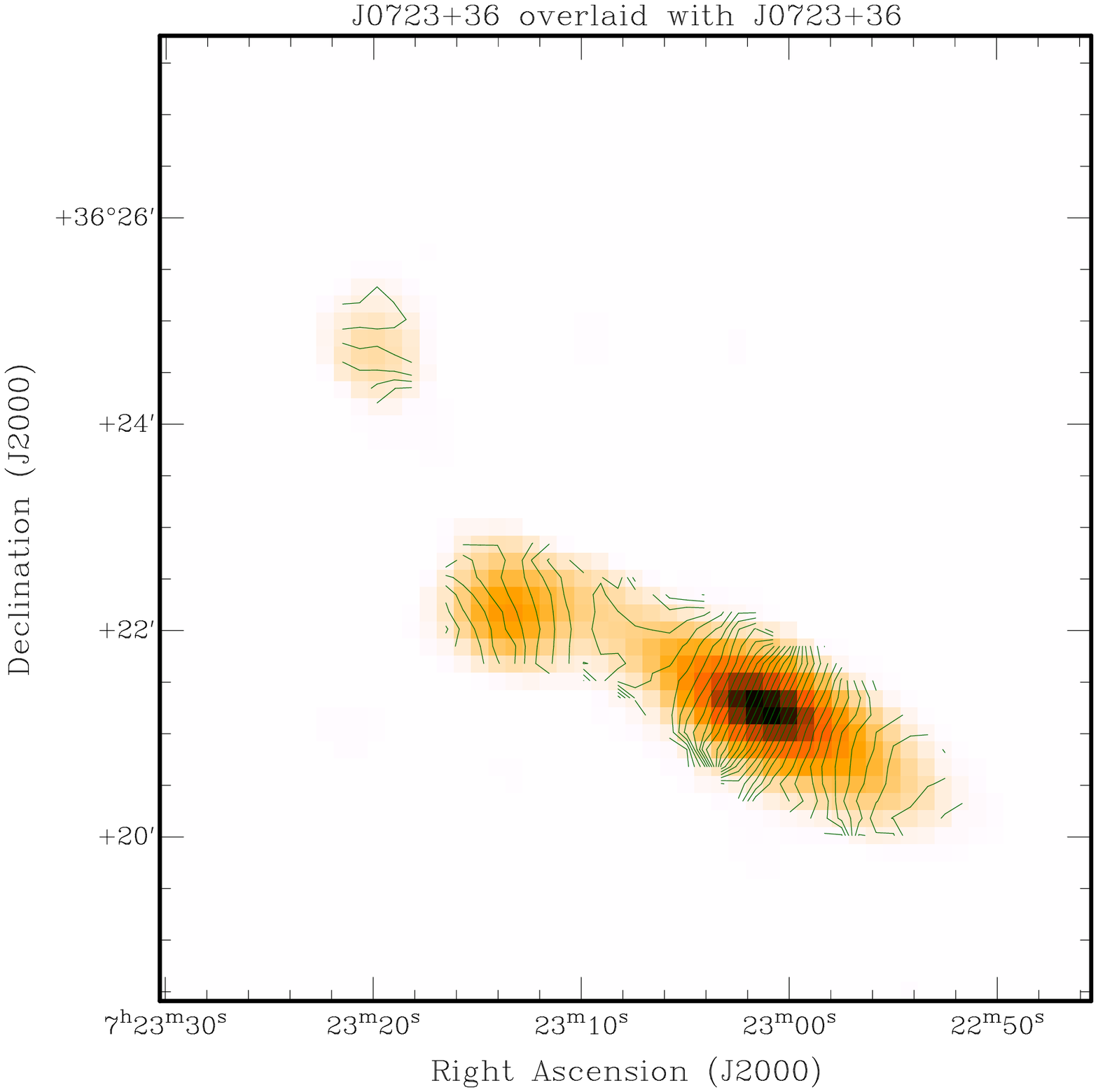}
 \includegraphics[angle=-0,width=6.5cm]{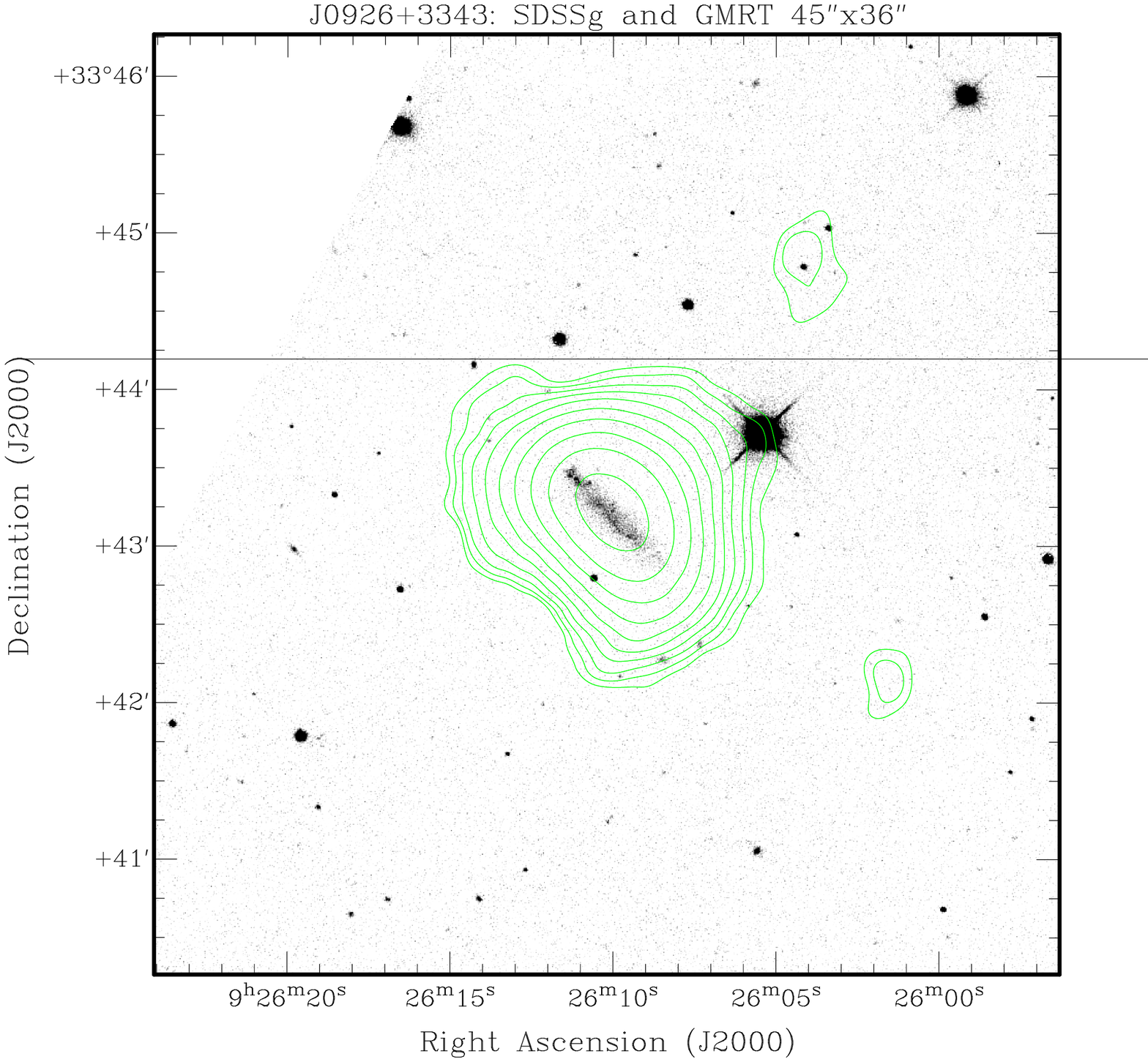}
 \caption{{\bf Left panel:}
Extremely gas-rich dwarf triplet J0723+36 near the centre of Lynx-Cancer
void, at $D$=16~Mpc, with $M$(HI)/$L_{\rm B}$ of $\sim$3, 10 and 25. The two
more massive
members appear to experience a minor merger, while the least massive and
most gas-rich dwarf at the NE is still well separated.
HI column density is shown in grey scale, while contures show isovelocity
lines with step of 6~km~s$^{-1}$. This finding can be a hint to possible
hidden void population of very gas-rich low mass galaxies (\cite{CP2013}).
{\bf Right panel:} Disturbed HI morphology in isolated void LSBD galaxies:
the 2-nd most metal-poor LSBD J0926+3343  (Chengalur et al., in prep.).
HI column density (in contures)
is superimposed on $g$-band SDSS image.  Evidence for cold accretion?
}
   \label{fig3}
\end{center}
\end{figure}

The third direction in study of void galaxies is related to their HI content
and structure. Integrated HI data on 96 Lynx-Cancer void galaxies (Pustilnik,
Martin, in prep.) indicate that void objects are in average gas-rich, with
median $M$(HI)/$L_{\rm B}$=1.2, $\sim$40\% higher than for the sample of
similar galaxies in denser environment. Mapping of their HI with Giant
Meterwave Radio Telescope leads to discovery of extremely gas-rich LSB dwarfs
with $M$(HI)/$L_{\rm B}$=10 and 25, and $M_{\rm gas}$/$M_{\rm bary} >$0.99
(\cite{CP2013}; see Fig.~\ref{fig3}, left). Such `unevolved' galaxies are
found mostly among the least luminous void galaxies. They can be not rare
among void objects with $M_{\rm B} \gtrsim$ --11, but due to severe
observational
selection effects they escape appearance in common wide-angle spectral
surveys.  Another interesting result of HI-mapping of three the most isolated
void LSBD galaxies shows their disturbed morphology (Fig.~\ref{fig3}, right).
Probably here we see the most clear cases of cold accretion along void
filaments (Chengalur et al. in prep.).

\section{Implications}

Summarising all above and some published findings on properties of galaxies
residing in nearby voids, we notice the following implications:
\begin{itemize}
\item
void galaxies in average show slower chemical evolution, having O/H in average
30\%--40\% lower than similar galaxies in denser environment; $\sim$10\%
of void galaxies have O/H lower by 2--5 times, indicating their unusual
evolutinory status.
\item
$ugri$ colours of outer parts  for $\sim$15\%  void galaxies
indicate the main SF episode started  $\sim$1--5 Gyr ago.
\item
More than a half of void
galaxies are gas-rich, with $M$(HI)/$L_{\rm B} >$1. Extremely gas-rich
dwarfs, with $M$(HI)/$L_{\rm B}$=4--25 already found in voids can be not
rare among the least luminous dwarfs ($M_{\rm B} >$--11).
\item
All together these results imply that evolution of void galaxies in
average goes substantially more slowly. In addition, there are indications
on that $\sim$10\% void galaxies formed with significant delay. This
fraction reaches
$\sim$30\% if we consider the least luminous void LSB dwarfs.
\end{itemize}

{\bf Acknowledgements.}
SAP is grateful to A.~Kniazev, J.-M.~Martin, ~J.Chengalur, A.~Tepliakova,
Y.~Perepelitsyna and E.~Safonova for fruitful collaboration and their
contribution to the discussed topics. The author acknowledges the partial
support of this work through RFBR grant 14-02-00520 and IAU travel grant.




\begin{thebibliography}{}

\bibitem[Abazajian et al. 2009]{DR7}
Abazajian K.N., Adelman-McCarthy J.K., Ag\"ueros M.A. et al.,
2009, ApJS, 182, 543

\bibitem[Berg et al.,  2012]{Berg12}
{Berg D.A.., Skillman E.D., Marble A.R., et al.} 2012,
\textit{ApJ}, 754, 98

\bibitem[Chengalur \& Pustilnik, 2013]{CP2013}
{Chengalur J.N., Pustilnik S.A.} 2013,
\textit{MNRAS} 428, 1579

\bibitem[Fairall, 1998]{Fairall98}
{Fairall A.,} 1998,
\textit{Large-Scale Structures in the Universe, Wiley-Praxis, 196 pp.}


\bibitem[Fioc \& Rocca-Volmerange, 1999)]{pegase2}
{Fioc M. \& Rocca-Volmerange B.} 1999,
\textit{arXiv:astro-ph/9912179}

\bibitem[Izotov \& Thuan (2007)]{IT07}
{Izotov Y.I., Thuan T.X.} 2007,
\textit{ApJ} 665, 1115

\bibitem[Kreckel et al., 2012)]{Kreckel12}
{Kreckel K., Platen E., Aragon-Calvo M.A., et al.} 2012,
\textit{AJ} 144, 16

\bibitem[Perepelitsyna et al., 2014)]{void_photo}
{Perepelitsyna Y.A., Pustilnik S.A., Kniazev A.Y.} 2014,
\textit{Astroph. Bull.} 69, 247 (arXiv:1408.0613)

\bibitem[Pustilnik \& Tepliakova (2011)]{PaperI}
{Pustilnik S.A., \& Tepliakova A.L.} 2011,
\textit{MNRAS} 415, 1188

\bibitem[Pustilnik et al. (2010)]{J0926}
{Pustilnik S.A., Tepliakova A.L., Kniazev A.Y., et al.}
2010, \textit{MNRAS} 401, 333

\bibitem[Pustilnik, Tepliakova \& Kniazev, 2011]{void_OH}
{Pustilnik S.A., Tepliakova A.L., Kniazev A.Y.} 2011,
\textit{Astroph. Bull.} 66, 255 (arXiv:1108.4850)

\bibitem[Pustilnik et al. (2011)]{void_LSBD}
{Pustilnik S.A., Tepliakova A.L., Kniazev A.Y.} 2011,
\textit{MNRAS} 417, 1335

\bibitem[Pustilnik et al. (2013)]{Eridanus}
{Pustilnik S.A., Martin J.-M., Lyamina Y.A., Kniazev A.Y.} 2013
\textit{MNRAS} 428, 1579

\end{thebibliography}
\end{document}